\documentclass[prd,twocolumn,floatfix,english,letterpaper]{revtex4}

\usepackage{amssymb}
\usepackage[dvips]{graphicx}

\usepackage[T1]{fontenc}
\usepackage[latin1]{inputenc}

\makeatletter

\providecommand{\tabularnewline}{\\}

\usepackage{babel}
\makeatother

\begin{document}

\title{Bottom baryons from a dynamical lattice QCD simulation}

\author{Randy Lewis}
\affiliation{Department of Physics and Astronomy, York University,
Toronto, Ontario, Canada M3J 1P3}

\author{R. M. Woloshyn}
\affiliation{TRIUMF, 4004 Wesbrook Mall, Vancouver,
British Columbia, Canada V6T 2A3}

\begin{abstract}
Bottom baryon masses are calculated based on a 2+1 flavor dynamical lattice QCD 
simulation. The gauge field configurations were computed by the CP-PACS and
JLQCD collaborations using an improved clover action. The bottom quark is 
described using lattice NRQCD. Results are presented for single and double-b
baryons at one lattice spacing. Comparison with experimental values is
discussed.

\end{abstract}

\maketitle

\section{\label{sec_intro}INTRODUCTION}

There have been a number of developments since our previous systematic
study \cite{NMnrqcd} of heavy baryons in quenched lattice QCD which 
suggest to us
to revisit this problem. On the experimental side, masses of four
bottom baryons have been measured since our earlier work. On the theory
side improved analysis methods should allow for a more precise determination 
of the masses and subsequently a more stringent test of the calculation.
As well, full dynamical simulations are becoming the norm. Dynamical
simulations for heavy baryons using lattice formulations different 
from that used in this work have been reported in Refs. \cite{Na} and
\cite{Burch}.
 
Besides lattice QCD, heavy baryons have been studied in many other
approaches. A few recent papers (from which the extensive earlier
literature may be traced) include work on QCD sum rules \cite{liu}, 
the quark model \cite{karlip,ebert} and the combined
heavy quark and $1/N_c$ expansions \cite{jenk}.

With a single b-quark and different flavor and spin combinations of
up, down and strange quarks eight different baryons can be constructed.
The properties of the baryons are summarized in Table \ref{tab_hlbary}. 
For the purposes
of this work exact isospin symmetry is assumed; u and d quarks are
taken to be degenerate. The experimental values for masses of 
$\Sigma_{b},\:\Sigma_{b}^{*},\:\Omega_{b}$
and $\Xi_{b}$ have only become available in the past 
year\cite{CDFsigma,D0omega,D0-CDFxi}. %
\begin{table}

\caption{Properties of single-b baryons showing valence content (q=u/d),
spin parity, isospin and mass (in GeV). The quantity $s_{l}$ is the total
spin of the light quark pair.}
\begin{center} 
\begin{tabular*}{0.45\textwidth}%
     {@{\extracolsep{\fill}}ccccccc}
\hline 
\hline 
Baryon&
quark content&
$J^{P}$&
I&
$s_{l}$&
mass&
Ref.\tabularnewline
\hline 
$\Lambda_{b}$&
qqb&
$\frac{1}{2}^{+}$&
0&
0&
5.624&\cite{pdg}
\tabularnewline
$\Sigma_{b}$&
qqb&
$\frac{1}{2}^{+}$&
1&
1&
5.812(3)&\cite{CDFsigma}
\tabularnewline
$\Sigma_{b}^{*}$&
qqb&
$\frac{3}{2}^{+}$&
1&
1&
5.8133(3)&\cite{CDFsigma}
\tabularnewline
$\Omega_{b}$&
ssb&
$\frac{1}{2}^{+}$&
0&
1&
6.165(16)&\cite{D0omega}
\tabularnewline
$\Omega_{b}^{*}$&
ssb&
$\frac{3}{2}^{+}$&
0&
1&
&
\tabularnewline
$\Xi_{b}$&
qsb&
$\frac{1}{2}^{+}$&
$\frac{1}{2}$&
0&
5.793(3)&\cite{D0-CDFxi}
\tabularnewline
$\Xi_{b}'$&
qsb&
$\frac{1}{2}^{+}$&
$\frac{1}{2}$&
1&
&
\tabularnewline
$\Xi_{b}^{*}$&
qsb&
$\frac{3}{2}^{+}$&
$\frac{1}{2}$&
1&
&
\tabularnewline
\hline
\hline
\end{tabular*}\end{center}
\label{tab_hlbary}
\end{table}

A variety of approaches are being pursued to incorporate dynamical
quark effects in lattice QCD simulations. Ideally one would like to
have u/d and strange (2+1 flavor) dynamical effects. As well, the
light quarks (u/d) should have small masses and the lattice volume
should be large. It is clear that these requirements stress the computing
resources available to even the largest lattice QCD collaborations
so some compromises have to be made. It is not feasible for us to
generate our own dynamical gauge field configurations. Instead 2+1
flavor dynamical configurations, made available by the CP-PACS and
JLQCD collaborations\cite{jlqcd1,jlqcd2}, were used. 
These are based on the Iwasaki RG
gauge field action\cite{iwasaki} and an 
improved clover action\cite{clover} for the quarks.

Lattice NRQCD\cite{latnrqcd} is used for the heavy quark. For the lattice spacing
considered here, only the b quark is heavy enough for its mass to
lie above the cut-off scale. Charm is too light to be simulated by
NRQCD and so, unlike Ref.~\cite{NMnrqcd}, charmed baryons are 
not considered in this study. 

The parameters in the calculation, namely u/d, s and b quark masses
as well as the overall scale, are fixed by calculations done in the
meson sector. Masses of heavy baryons are then predictions of the
simulation. In addition to single-b baryons listed in Table \ref{tab_hlbary}
we also
calculate the masses of the double-b baryons with properties given
in Table \ref{tab_hhbary}. 
At present there are no data for double-b baryons but
it is hoped that eventually they will be observed in some future experiment.

\begin{table}

\caption{Properties of double-b baryons showing valence content (q=u/d),
spin parity and isospin.}
\begin{center}
\begin{tabular*}{0.45\textwidth}%
     {@{\extracolsep{\fill}}cccc}
\hline 
\hline 
Baryon&
quark content&
$J^{P}$&
I\tabularnewline
\hline 
$\Xi_{bb}$&
qbb&
$\frac{1}{2}^{+}$&
$\frac{1}{2}$\tabularnewline
$\Xi_{bb}^{*}$&
qbb&
$\frac{3}{2}^{+}$&
$\frac{1}{2}$\tabularnewline
$\Omega_{bb}$&
sbb&
$\frac{1}{2}^{+}$&
0\tabularnewline
$\Omega_{bb}^{*}$&
sbb&
$\frac{3}{2}^{+}$&
0\tabularnewline
\hline 
\hline
\end{tabular*}
\end{center}
\label{tab_hhbary}
\end{table}

In Sec.~\ref{sec_sim} details of the simulation are presented. There is a brief
summary of the actions and the lattice parameters. The analysis method
is also discussed.

Results are given in Sec. \ref{sec_res} and compared to available experimental
data. Limitations of this work and future directions are discussed
in Sec. \ref{sec_summ}.

\section{\label{sec_sim}NUMERICAL SIMULATION}

\subsection{\label{sec_detail}Simulation details}

Gauge field configurations incorporating the full dynamical vacuum
polarization effects of u,d and s quarks are employed in this work.
These gauge configurations were generated by the CP-PACS and JLQCD
collaborations\cite{jlqcd1,jlqcd2}
and made available through the Japan Lattice Data Grid\cite{jldg}.

The gauge field action is an improved action developed by Iwasaki\cite{iwasaki}
and in addition to the standard plaquette term it contains six-link
operators with coefficients tuned so that the action lies close to
the renormalized trajectory. The quark action is of the clover 
type\cite{clover}.
The coefficient of the clover was tuned nonperturbatively\cite{clovercoeff}
and is the
same for all values of sea-quark mass. Hybrid Monte Carlo was used
for the two light flavors (u/d) and the Polynomial Hybrid Monte Carlo
was used for the strange sea quark\cite{phmc}. 

In this work, a subset of the CP-PACS/JLQCD 2+1 flavor configurations
at $\beta=$ 1.90 with lattice size $20^{3}\times40$ is used. The
coefficient of the clover term in the quark action was 1.715 
(see Ref. \cite{jlqcd2}). 
Other lattice parameters are given in Table \ref{tab_latparms}. 
The quantity $U_{0}$ equals $1/3$ of the trace of the mean 
gauge-field link in Landau gauge
and is used as the tadpole factor in the NRQCD action.

\begin{table}

\caption{Lattice parameters. The lattice size is $20^{3}\times40$ with Iwasaki
gauge action $\beta=$ 1.90 and quark action clover coefficient $c_{SW}=$
1.715. The masses of u/d and s quarks are encoded in the hopping parameters
$\kappa_{q}$ and $\kappa_{s}$ respectively.}

\begin{center}
\begin{tabular*}{0.45\textwidth}%
     {@{\extracolsep{\fill}}cccc}
\hline 
\hline 
configurations&
$\kappa_{q}$&
$\kappa_{s}$&
$U_{0}$\tabularnewline
\hline 
480&
0.1358&
0.1358&
0.8396\tabularnewline
576&
0.1364&
0.1358&
0.8415\tabularnewline
576&
0.1368&
0.1358&
0.8425\tabularnewline
576&
0.1370&
0.1358&
0.8432\tabularnewline
480&
0.1358&
0.1364&
0.8405\tabularnewline
576&
0.1364&
0.1364&
0.8422\tabularnewline
576&
0.1368&
0.1364&
0.8433\tabularnewline
576&
0.1370&
0.1364&
0.8439\tabularnewline
\hline
\hline
\end{tabular*}\end{center}
\label{tab_latparms}
\end{table}

A non-relativistic action\cite{latnrqcd} is used to describe the b quark. 
This approach
is useful when the bare mass of the quark is larger than the cutoff
scale. The particular form of the action used in this work is essentially
the same as that used in our previous works\cite{NMnrqcd,orderM3}. 
Terms up to $\mathcal{O}$$(1/M_{0}^{3})$
in the heavy quark mass $M_{0}$ are retained. The only difference
is that while \cite{NMnrqcd,orderM3} used an anisotropic lattice, 
here the lattice is
isotropic. The details of the action are given in the Appendix.

Simulations were done for three values of the heavy-quark bare mass:
2.28, 2.34 and 2.40 in lattice units. This allowed for interpolation
to the physical b-mass value.

Correlation functions were calculated using local hadron operators
at the source and sink. For heavy baryons the operators used were
exactly the same as detailed in Ref.~\cite{NMnrqcd}. In addition to baryon 
correlators, correlators were calculated for pseudoscalar and vector mesons in
the light sector (u/d,s), the heavy-light sector (B-mesons) and the
heavy-heavy sector ($\Upsilon$). The meson calculations were used
to determine the ``physical'' values for the quark masses and
the overall scale. As well they provide some additional predictions
which test the calculation.

Non-relativistic quarks propagate only forward in time. For combining
relativistic quarks with non-relativistic quarks it is convenient to
use Dirichlet boundary conditions for the light quark propagators;
propagation across the time boundary is not allowed. If the source time is
set in from the time boundary (four time steps is used in this calculation)
it can be verified that meson masses in the light sector are the same,
within statistical errors, as those computed with more usual periodic
time boundary conditions.

In order to get some extra suppression of statistical fluctuations
and reach the level of precision that we would like, multiple sources were used.
For each gauge field configuration a set of correlators was calculated
using different space-time points as the source point. This set was
averaged and the average value was used as the representative correlation
function for the configuration. Eight sources per configuration were
used in this calculation.

\subsection{\label{sec_anal}Analysis}

Correlation functions were fit with a sum of exponentials
\begin{equation}
g(t)=\sum_{i=0}^{n-1}z_{i}e^{-E_{i}t}
\end{equation}
over a fixed time range which for our standard analysis extended
for 27 time steps starting one time step past the source position.
For mesons three exponential terms were found to be adequate while
for baryons four terms were used. The advantage of a multi-exponential
fit over methods such as fitting over an {}``effective mass'' plateau
is that it is less subjective. Also it makes better use of the correlation
function data at times where the statistical error is small. 

To stabilize the fits a constrained fitting method\cite{constrainedfit} 
was used. The usual
$\chi^{2}$ is augmented with a term which acts to prevent the fit
parameters from straying outside some sensible (but broad) range.
The constraint term for the coefficients $z_{i}$ was taken to be
\begin{equation}
\sum_{i=0}^{n-1}\frac{(z_{i}-\bar{z}_{i})^{2}}{\sigma_{\bar{z}_{i}}^{2}}.
\end{equation}
The priors were chosen to be $\bar{z}_{i}=\sigma_{\bar{z_{i}}}=g(0)/n$,
which is a very loose constraint. The constraint term for the exponents
(energies) has a similar form 
\begin{equation}
\sum_{i=0}^{n-1}\frac{(E_{i}-\bar{E}_{i})^{2}}{\sigma_{\bar{E}{}_{i}}^{2}}.
\end{equation}
One might be tempted to choose a constraint which is minimally biased
such as equal spacing of the energies $\bar{E}_{i}-\bar{E}_{i-1}=\delta E$
and $\sigma_{\bar{E}_{i}}=\delta E.$ In practice it was found that
to prevent the occurrence of obviously spurious solutions, where for
example two terms have the same exponent and coefficients of opposite
sign, a somewhat tighter constraint is needed. For our fits we use
$\bar{E}_{1}-\bar{E}_{0}=0.5$ and $\bar{E}_{i}-\bar{E}_{i-1}=1.0,i>1$
with $\bar{E}_{0}$ equal to $0.5$ for mesons and $0.8$ for baryons.
The $\sigma_{\bar{E}_{i}}$ were chosen to be proportional to the
spacing between the $\bar{E}_{i}$ with a common reduction factor
taken to be 0.83. With this setup spurious solutions were avoided
and the ground state energy was very stable with respect to changes
in the constraint parameters. For example, it was found that changing 
prior parameters by 20\% led to changes in single-b baryon simulation
energies (which are most sensitive) of about 1\%. This effect is
incorporated into the estimated systematic uncertainty.

To determine the statistical errors, bootstrap analysis was used.
A sample of 600 bootstrap ensembles was created and a complete analysis
was carried out for each ensemble. In this way the uncertainty in
determining the quark mass parameters and the scale is incorporated
into the bootstrap error estimate of the final result.

A few systematic effects were examined explicitly. These include sensitivity
to the choice of time range for fitting the correlators, to the determination
of the b-quark mass and to the choice of extrapolation function in
u/d and s quark mass. In addition, there are systematic errors associated
with the omitted higher order effects in the NRQCD action. These are
discussed in great detail by Gray {\emph {et al.}}\cite{gray} for the 
$\Upsilon$ system. For heavy-light hadrons the appropriate power
counting is different than for quarkonium\cite{manohar}. 
The expansion parameter is $p/M_0$ where the typical momentum of the
heavy quark $p\sim\Lambda_{QCD}$, independent of quark mass. Radiative
corrections to the action then are $\mathcal{O}(\alpha_s\Lambda_{QCD}/M_0)$
relative to the leading kinetic term\cite{manohar} which we estimate 
can induce a 1.5\% relative uncertainty in the simulation 
energies. The NRQCD action is corrected for $\mathcal{O}(a^2)$ 
lattice spacing errors at tree level. The leading discretization 
corrections are $\mathcal{O}(\alpha_s(a\Lambda_{QCD})^2)$ and a 0.5\%
fractional systematic error is assigned to these effects. The action 
used in this work includes $\mathcal{O}(1/M^3_0)$ terms and higher 
order relativistic effects are assumed to be negligible.

\section{\label{sec_res}RESULTS}

\subsection{\label{sec_mes}Mesons}

The first task is to determine the quark mass parameters and overall
scale. We start with the u/d and s sector and use the pion, rho meson
and phi meson masses as experimental input. Lattice masses for these
mesons are calculated with the eight ensembles in Table III. For each 
correlator 
the valence mass was taken to be equal to the u/d mass or to
the strange mass. No partially quenched correlators, with a valence mass
different from a sea-quark mass of the same flavor were used in this
work.

The lattice
meson masses were then fit as a function of quark mass using the vector
Ward identity (VWI) definition of the quark mass $m=(1/\kappa-1/\kappa_{cr})/2$
where $\kappa_{cr}$ is the point where the pseudoscalar meson mass
vanishes when all valence and sea quark masses have this hopping parameter.

The fitting functions are motivated by the study of light quark masses
in Ref.~\cite{jlqcd2}. For the pseudoscalar meson (with degenerate 
valence quarks) it was found that the three terms 
\begin{equation}
b_{1}m_{v}+b_{2}(2m_{q}+m_{s})+b_{3}m_{v}(2m_{q}+m_{s}),
\label{eq_psfit}
\end{equation}
where $m_{v}$, $m_{q}$ and $m_{s}$ are valence, u/d and s quark masses
from  the VWI definition, gives a good description of the mass squared
lattice data. Note that $\kappa_{cr}$ is a free parameter in the
fit as are the coefficients $b_{1},b_{2},b_{3}.$ 

For the vector meson the mass was fit using
\begin{equation}
c_{1}+c_{2}m_{v}+c_{3}(2m_{q}+m_{s})+c_{4}m_{v}^{2}
\label{eq_quadfit}
\end{equation}
 which takes into account a slight nonlinear dependence on the quark
mass. In determining the coefficients of this fit $\kappa_{cr}$ is
fixed to the value obtained in the pseudoscalar meson fit.

After the fit parameters are determined one can solve for the values
of the hopping parameters and the lattice spacing
that reproduce the input experimental numbers.
The results are $\kappa_{q}=0.13784(2)$ and $\kappa_{s}=0.13618(9)$
with a value of the inverse lattice spacing $a^{-1}=1.89(3)$GeV.
We note that the inverse lattice spacing obtained here is slightly different
from that quoted in \cite{jlqcd2} but it has to be remembered that the
number of configurations, the number of ensembles and the details of the
mass extrapolation are also different.

The b-quark mass is determined using the $\Upsilon$ mass as input.
In NRQCD the quark mass has been removed from the action. The zero-momentum
hadron correlator yields a simulation energy to which the renormalized
quark mass and energy shift must be added to get the hadron mass. Alternatively
the hadron mass can be determined from the kinetic energy. For this
purpose the correlators for $b\overline{b}$ mesons carrying a unit
of momentum were calculated. The meson mass M (lattice units) is obtained using
\begin{equation}
M=\frac{2\pi^{2}}{N_{s}^{2}(E_{sim}(p)-E_{sim}(0))}
\end{equation}
 where $N_{s}$ is the spatial extent of the lattice, $p=2\pi/N_{s}$
and the difference in simulation energies is just the kinetic energy.

Kinetic energies were calculated for $\Upsilon$ at all the heavy
bare quark masses listed in Sec.~\ref{sec_detail} and for all eight ensembles. It
was found that kinetic energies were practically independent of the
values of the sea quark masses with no systematic trend which would
allow for extrapolation. Therefore for each ensemble an independent
determination of the bare b-quark mass was made by inputting the experimental
$\Upsilon$ mass and the inverse lattice spacing obtained above. The
values (in lattice units) varied in the small range from 2.375(78)
to 2.416(77). A value in the middle of this range 2.391(78), obtained
from the ensemble $\kappa_{q}=0.1368,\kappa_{s}=0.1364$, was
used as our nominal value for the b mass. The maximum and minimum values
were used to estimate a systematic uncertainty in the hadron masses.
The effect of choosing different values for the b-quark mass was found to be
small, typically changing the final hadron mass by an MeV or so.

Having determined the physical point for the b-quark mass one can
get the splitting between the $\Upsilon$ and the
$\eta_{b}$ just from interpolating the difference in simulation energies
for the $b\bar{b}$ vector and pseudoscalar channels.
The calculated  $\Upsilon-\eta_{b}$ mass difference is given in Table IV
and is smaller than
the recent value reported by the BaBar collaboration\cite{BabarEtab}. 
The underestimate of the quarkonium spin splitting
has been a common feature of simulations
done with NRQCD\cite{trottier,collins,koniuk}. The systematic study by
Gray {\emph {et al.}}\cite{gray} (see their Fig. 14) shows that this quantity is 
quite sensitive to the continuum extrapolation for light sea-quark masses. 
Gray {\emph {et al.}}\cite{gray}
predicted a continuum extrapolated value of 61(14)MeV, somewhat larger than our
value at a non-zero lattice spacing. Note that part of the
difference between their value and ours is due to $\mathcal{O}(v^6)$ terms in 
the NRQCD action which we include and they do not. From a test run on
a single ensemble it is estimated that these terms 
decrease the spin splitting by about 15\%. 

\begin{table}

\caption{Lattice results for meson masses and spin splittings (in GeV) compared
to experimental values. The $\Upsilon-\eta_{b}$ result is from
\cite{BabarEtab}, other experimental values are from \cite{pdg}.}
\begin{center}
\begin{tabular*}{0.45\textwidth}%
     {@{\extracolsep{\fill}}cccc}
\hline 
\hline 
&mass-$M_{\Upsilon}/2$&
mass&
Experiment\tabularnewline
\hline 
$\Upsilon-\eta_{b}$&&
0.039(1)$(_{7}^{8})$&
0.0714($_{23}^{31})$(27)
\tabularnewline
$B$&0.527(6)(8)&
5.257(6)(8)&
5.2793(4)\tabularnewline
$B^{*}$&0.530(7)($_{9}^{10})$&
5.300(7)($_{9}^{10})$&
5.325(1)\tabularnewline
$B_{s}$&0.617(3)(10)&
5.346(3)(10)&
5.366(1)\tabularnewline
$B_{s}^{*}$&0.658(4)(11)&
5.388(4)(11)&
5.412(1)\tabularnewline
$B^{*}-B$&&
0.043(3)($_{4}^{5}$)&
0.0458(4)\tabularnewline
$B_{s}^{*}-B_{s}$&&
0.042(2)($_{4}^{5})$&
0.0461(15)\tabularnewline
\hline
\hline
\end{tabular*}\end{center}
\label{tab_mesonmasses}
\end{table}

Although the main motivation for this work was to study heavy baryons,
the masses for heavy-light (B) mesons were also calculated. The masses
were computed relative to the $\Upsilon$ which was an input to the
calculation. Above, the kinetic energy was used to determine the $\Upsilon$
mass but alternatively it is given as $M_{\Upsilon}=E_{sim}^{(\Upsilon)}+2(ZM_{0}-E_{shift})$
where $Z$ is the mass renormalization factor and $E_{shift}$ is
an additive mass shift. The quantities $Z$ and $E_{shift}$ are independent
of the hadronic state\cite{davies} so for a hadron, such as the B-meson, containing
a single b quark $M_{B}=E_{sim}^{(B)}+ZM_{0}-E_{shift}.$ The mass
can then be obtained using 
\begin{equation}
M_{B}=E_{sim}^{(B)}+\frac{1}{2}(M_{\Upsilon}-E_{sim}^{(\Upsilon)}).
\label{eq_bmass}
\end{equation}
 The results, extrapolated in light quark mass and converted to physical
units, are shown in Table \ref{tab_mesonmasses} along with the experimental 
values from the PDG\cite{pdg}. The first error in the lattice results is 
the statistical (bootstrap) error. The second error incorporates the changes 
that result from changes in the time range used in fitting the correlators,
the uncertainty in determining the b-quark mass and the choice of
u/d and s quark mass fitting function. For the mass range used in
this present simulation the heavy-light meson masses are consistent
with a linear dependence on quark mass
\begin{equation}
c_{1}+c_{2}m_{v}+c_{3}(2m_{q}+m_{s})
\label{eq_3parmfit}
\end{equation}
 where $m_{v},m_{q}$ and $m_{s}$ are the VWI quark masses. This
was the form used to extrapolate for u/d and interpolate for s. The
systematic error incorporates an estimate of sensitivity to non-linear
quark mass dependence made by adding a quadratic term as in Eq.
(\ref{eq_quadfit}). Also included in the systematic error are estimated
uncertainties associated with radiative and discretization corrections 
to the NRQCD action.

The heavy-light meson results look quite reasonable although there
is a systematic underestimate of the mass by about 25MeV compared 
to experimental values. It is plausible that a common effect underlies 
this trend. At this stage,
we don't know the effect of changing the lattice spacing
so it is natural to suspect that the overall discrepancy in 
Table \ref{tab_mesonmasses} is due to a small uncorrected lattice
spacing error. Assuming the lattice spacing error associated with
the light quark action is $\mathcal{O}(a^2\Lambda^2_{QCD})$ one gets the
estimate $a^2\Lambda^2_{QCD}$(mass-$M_{\Upsilon}/2$) $\sim$ 0.015GeV
which is roughly the size of the observed discrepancy.
Note that for heavy-light mesons, where the 
wavefunctions are not so strongly affected by short distance interactions
as in bottomonium, there is good agreement of the calculated 
spin splittings with experimental values.

\subsection{\label{sec_bar}Baryons}

\begin{figure}
\scalebox{0.55}{\includegraphics*{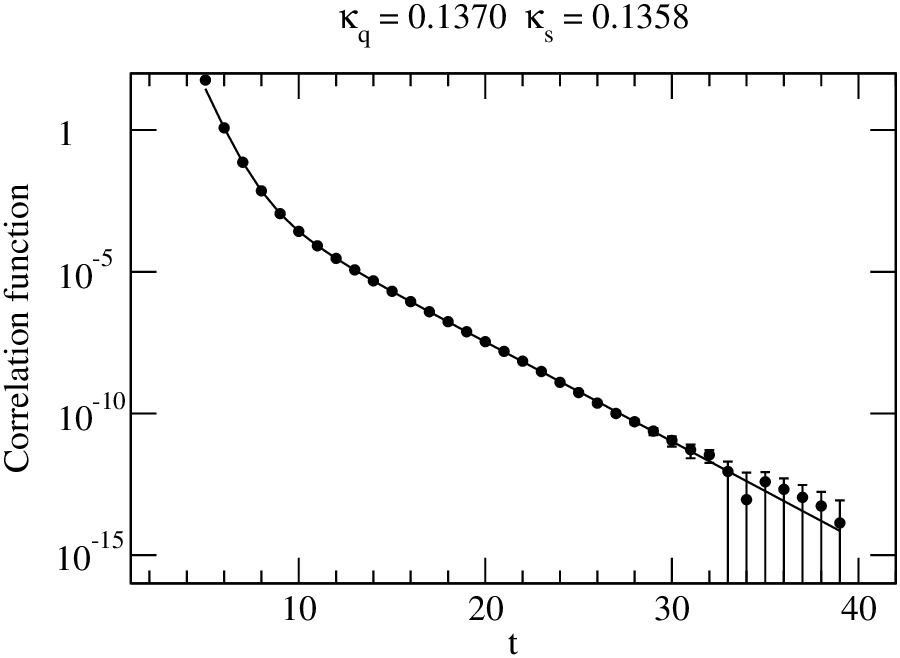}}
\caption{A $\Sigma_{b}$ correlator as a function of lattice time. The lattice
boundaries are at t equals 1 and 40. The source is at t equals 5.
The fit (solid line) is done including points t equals 6 to 32.}
\label{sigmacf}
\end{figure}

The calculation of baryon masses proceeds very much like that for
heavy-light mesons. The simulation energies were determined by doing
four-term exponential fits to the baryon correlation function. A typical
example, showing the quality of the simulation data and of the fit,
is given in Fig.~\ref{sigmacf}. For single-b baryons, masses were calculated
using the baryon analog of Eq.~(\ref{eq_bmass}). The resulting masses were 
first interpolated in the b-quark mass to the physical point determined
by $M_{\Upsilon}$. Then the u/d and s quark mass dependence was fit
using Eq. (\ref{eq_3parmfit}). A linear quark mass dependence was 
consistent with all simulation data. A quadratic valence mass dependence 
was used to estimate a systematic uncertainty.

\begin{table}

\caption{Lattice results for masses of single-b baryons (in GeV) compared
to experimental values.}

\begin{center} 
\begin{tabular*}{0.45\textwidth}%
     {@{\extracolsep{\fill}}ccccc}
\hline 
\hline 
&
mass-$M_{\Upsilon}/2$&
mass&
Experiment&
Ref.\tabularnewline
\hline 
$\Lambda_{b}$&
0.911(21)($_{33}^{15}$)&
5.641(21)($_{33}^{15}$)&
5.620(2)&\cite{pdg}
\tabularnewline
$\Sigma_{b}$&
1.065(16)($_{26}^{17}$)&
5.795(16)($_{26}^{17}$)&
5.8115(30)&\cite{CDFsigma}
\tabularnewline
$\Sigma_{b}^{*}$&
1.112(26)($_{18}^{20}$)&
5.842(26)($_{18}^{20}$)&
5.8327(34)&\cite{CDFsigma}
\tabularnewline
$\Omega_{b}$&
1.276(10)($_{19}^{20}$)&
6.006(10)($_{19}^{20}$)&
6.165(16)&\cite{D0omega}
\tabularnewline
$\Omega_{b}^{*}$&
1.314(18)($_{21}^{20}$)&
6.044(18)($_{21}^{20}$)&
&
\tabularnewline
$\Xi_{b}$&
1.051(17)($_{16}^{17}$)&
5.781(17)($_{16}^{17}$)&
5.7929(30)&\cite{D0-CDFxi}
\tabularnewline
$\Xi'_{b}$&
1.173(12)($_{19}^{18}$)&
5.903(12)($_{19}^{18}$)&
&
\tabularnewline
$\Xi_{b}^{*}$&
1.220(21)($_{21}^{19}$)&
5.950(21)($_{21}^{19}$)&
&
\tabularnewline
\hline
\hline
\end{tabular*}\end{center}
\label{tab_hlmasses}
\end{table}

\begin{figure}
\scalebox{0.55}{\includegraphics*{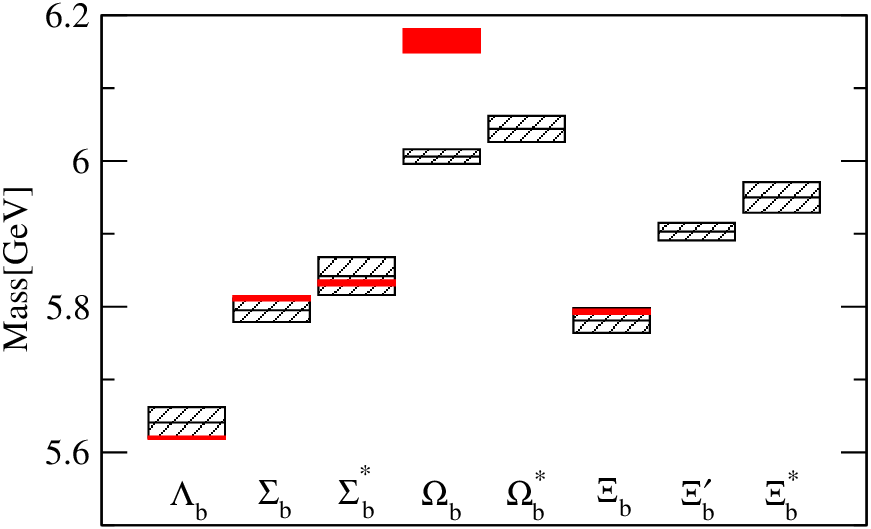}}
\caption{Masses of single-b baryons. The diagonally-hatched boxes are lattice
results with statistical errors only. Solid bars (red) are experimental 
values.}
\label{hlmassstat}
\end{figure}

\begin{figure}
\scalebox{0.55}{\includegraphics*{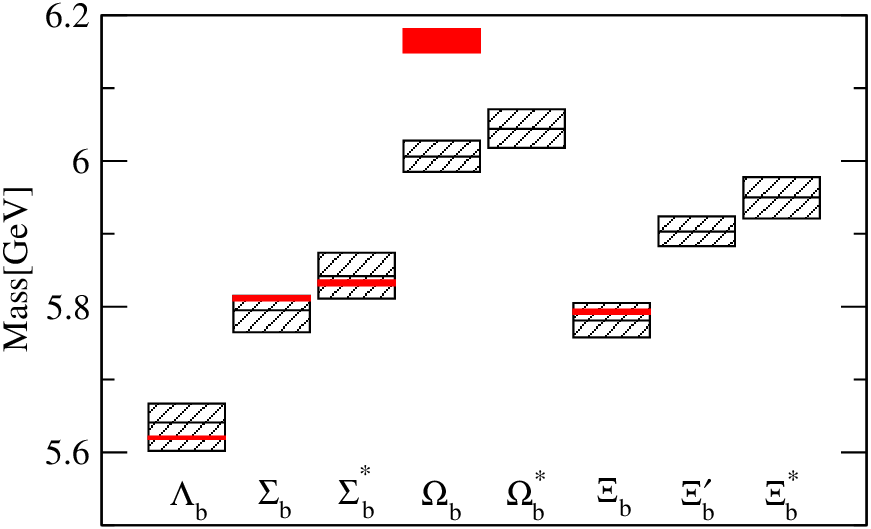}}
\caption{Masses of single-b baryons. The diagonally-hatched boxes are lattice
results with combined statistical and systematic errors. 
Solid bars (red) are experimental values.}
\label{hlmasscomb}
\end{figure}

The results for single-b baryons are tabulated in Table \ref{tab_hlmasses} 
and plotted
in Fig.~\ref{hlmassstat} with statistical errors and in 
Fig.~\ref{hlmasscomb} with combined statistical
and systematic errors. The $\Lambda_{b}$ and $\Sigma_{b}$
show significant sensitivity
to the inclusion of a quadratic term in the mass fit; other masses
are not changed very much by this term. 
The present results show a vast improvement in statistical
precision compared to our previous study\cite{NMnrqcd}.
Multi-exponential
constrained fitting which uses time-correlation information over a
large range, including times where the correlation function statistical errors
are small, plays an important role in this improvement. Also shown in 
Table \ref{tab_hlmasses} is the quantity mass-$M_{\Upsilon}/2$. This is the
actual quantity that the lattice simulation provides and, since most of the 
heavy hadron mass is due to the b-quark mass, it is a rough measure of the
quark and gluon interaction energy. With the present analysis methods this 
quantity is determined to a few percent.

With the exception of $\Omega_{b}$ the results are in good agreement 
with the experimental values. However, one should not
overinterpret this agreement since scaling (with lattice spacing) has not
been checked and effects 
of using more realistic u/d quark masses still have to be considered.

\begin{figure}
\scalebox{0.55}{\includegraphics*{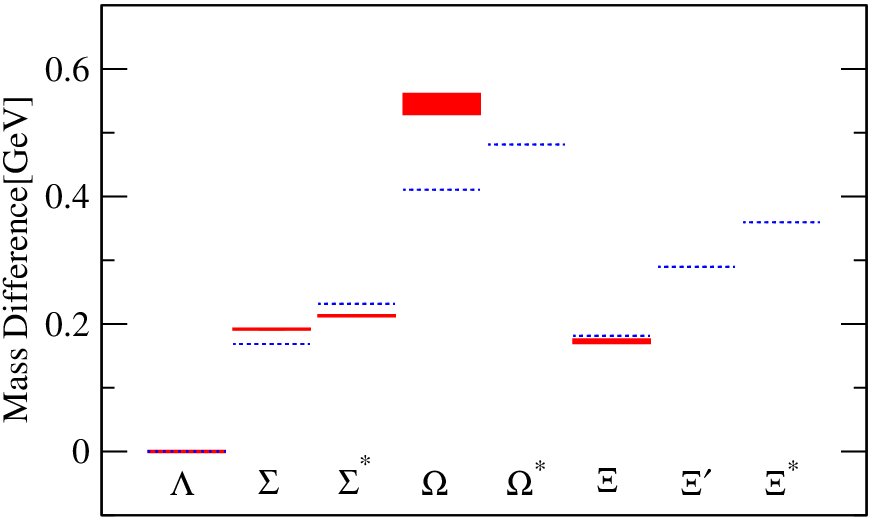}}
\caption{Experimentally measured values of masses of 
single-charm (dashed, blue)
and single-bottom baryons (solid, red) relative to the lowest lying
state $\Lambda$.}
\label{hlmassDiff2}
\end{figure}

\begin{figure}
\scalebox{0.55}{\includegraphics*{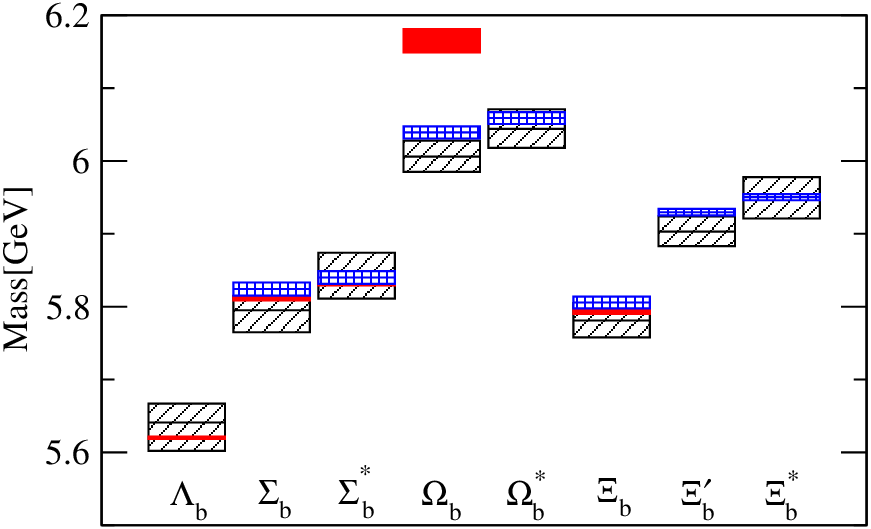}}
\caption{Masses of single-b baryons from Jenkins\cite{jenk} 
(vertically-hatched boxes, blue). The diagonally-hatched boxes are 
lattice results and solid bars (red) are experimental values.}
\label{hlmassJenk2}
\end{figure}

The large discrepancy between our calculated $\Omega_{b}$ mass and the value
reported in \cite{D0omega} is perplexing. To understand how puzzling it is, 
consider the basic physical idea behind heavy quark effective theory: a heavy
quark acts essentially as a static color source and so mass differences 
between states with different light-quark configurations should be
independent of heavy quark mass (up to corrections inversely proportional
to the heavy mass). This suggests a direct data-to-data comparison of
mass differences in single-charm and single-bottom baryons as shown in
Fig. \ref{hlmassDiff2}. The measured masses of the singly-heavy 
$\Sigma, \Sigma^{*}$ and $\Xi$ baryons fit the expected pattern but  
$\Omega$ shows a large discrepancy, the same behavior as observed with
our lattice simulation. The application of heavy quark effective theory to
singly-heavy baryons was formalized by Jenkins\cite{jenk2}. In this work a 
combined expansion in the inverse of the heavy-quark mass, in $1/N_c$ and 
in SU(3) flavor symmetry breaking was carried out. Mass formulas were
derived which then allow some masses to be predicted in terms of other
experimentally measured masses. This is a more rigorous version of our
data-to-data comparison. The updated predictions for single-b baryons
from Jenkins\cite{jenk} are shown in Fig. \ref{hlmassJenk2} along with 
our lattice results and the experimental values. There is consistency 
between our lattice results and the effective theory analysis of 
\cite{jenk,jenk2} across the whole spectrum which further accentuates 
the $\Omega_{b}$ puzzle.

\begin{table}

\caption{Lattice results for masses of double-b baryons (in GeV).}

\begin{center}
\begin{tabular*}{0.40\textwidth}%
     {@{\extracolsep{\fill}}ccc}
\hline 
\hline 
&
mass-$M_{\Upsilon}$&
mass\tabularnewline
\hline
$\Xi_{bb}$&
0.667(13)($_{26}^{12}$)&
10.127(13)($_{26}^{12}$)\tabularnewline
$\Xi_{bb}^{*}$&
0.691(14)($_{25}^{16}$)&
10.151(14)($_{25}^{16}$)\tabularnewline
$\Omega_{bb}$&
0.762(9)($_{13}^{12}$)&
10.225(9)($_{13}^{12}$)\tabularnewline
$\Omega_{bb}^{*}$&
0.786(10)($_{12}^{18}$)&
10.246(10)($_{12}^{18}$)\tabularnewline
$\Xi_{bb}^{*}-\Xi_{bb}$&
&
0.026(8)($_{10}^{11}$)\tabularnewline
$\Omega_{bb}^{*}-\Omega_{bb}$&
&
0.025(7)($_{6}^{11}$)\tabularnewline
\hline
\hline
\end{tabular*}\end{center}
\label{tab_hhmasses}
\end{table}

\begin{figure}
\scalebox{0.55}{\includegraphics*{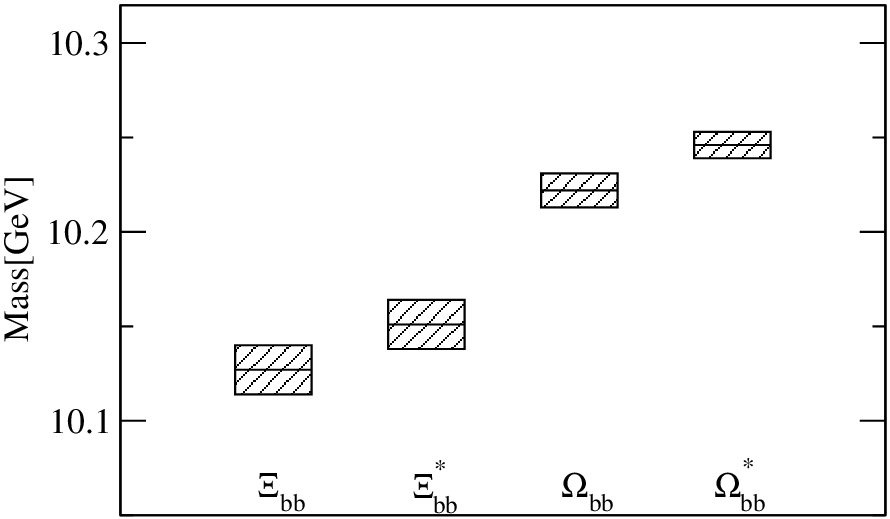}}
\caption{Lattice results for masses of double-b baryons showing
statistical errors only.}
\label{hhmassstat}
\end{figure}

\begin{figure}
\scalebox{0.55}{\includegraphics*{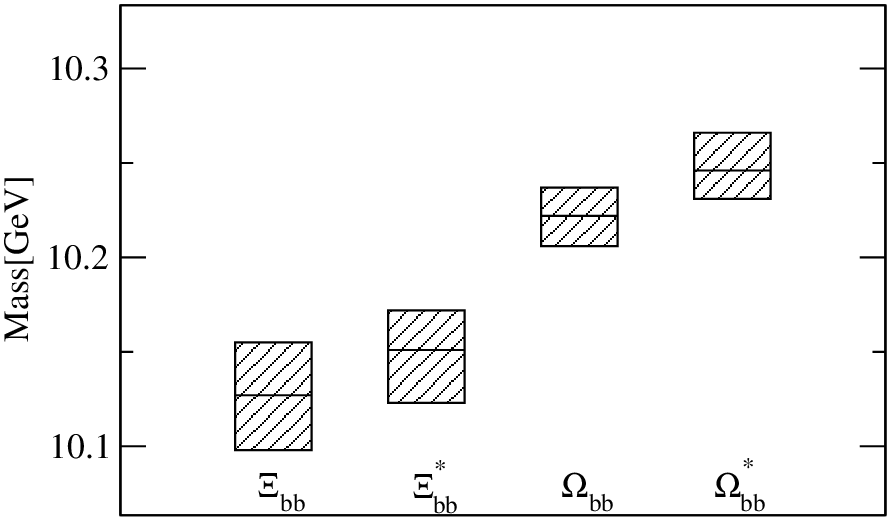}}
\caption{Lattice results for masses of double-b baryons showing
combined statistical and systematic errors.}
\label{hhmasscomb}
\end{figure}

The masses of double-b baryons are calculated using
\begin{equation}
M_{bb}=E_{sim}^{(bb)}+M_{\Upsilon}-E_{sim}^{(\Upsilon)}.
\end{equation}
The masses are interpolated and extrapolated as above. The final
results are listed in Table \ref{tab_hhmasses}
and shown in Figs.~\ref{hhmassstat} and \ref{hhmasscomb} with statistical
and combined errors respectively. As yet there are no experimental
values to compare with these calculations. The mass difference between
spin 3/2 and spin 1/2 states in the doubly heavy sector is interesting
because in the quark model and in heavy quark effective theory the
baryon spin splitting can be related to the spin splitting for 
heavy-light mesons\cite{lipkin,savwise,charm,others}. 
Simple arguments suggest that 
\begin{equation}
\Delta M_{baryon}\approx\frac{3}{4}\Delta M_{meson}.
\end{equation}
Within errors, our lattice spin splittings are consistent with this relation 
but are not precise enough to test it very stringently.

\section{\label{sec_summ}SUMMARY AND DISCUSSION}

In this paper the masses of bottom baryons were calculated in a 2+1
flavor dynamical lattice simulation. The clover action was used for
light (u/d and s) quarks and the b quark was described by NRQCD. Using
constrained multi-exponential fitting for hadron correlators led to
much more precise results than our previous studies. Single bottom
baryons whose masses can be compared to experimental values show good
agreement with the exception of the recently measured $\Omega_b$.

The $\Omega_b$ is puzzling since our lattice result is consistent
with ideas based on heavy quark effective theory which allow for
a prediction of the $\Omega_b$ mass using only empirical input. 

Predictions are made for the still unobserved single-b
baryons and for the double-b baryons. Experimental observation of any
of these states would be extremely interesting and might shed some 
light on the $\Omega_b$ puzzle.

This study should be considered as the first step in a program to
do precision calculations in the heavy baryon sector. To go further,
some systematic improvements have to be made. Radiative corrections
to the NRQCD action are one of the major contributions to the systematic
error and have to be dealt with. 
The light quark action is not corrected for lattice spacing errors
to the same extent as the heavy quark action. Calculations are needed
at more than one lattice spacing to enable a continuum extrapolation
and to estimate reliably lattice spacing errors.

The present simulation is done in a region where the u/d quark masses are 
about 0.4 times the strange mass and some baryon masses 
exhibit a sensitivity to light quark mass extrapolation. 
Smaller values for the u/d mass would be highly desirable to
insure that the simulation captures more accurately the dynamical
sea quark effects and that the extrapolation to the physical point
can be put on a firm theoretical basis. Using an array of algorithmic 
improvements, the PACS-CS collaboration has produced configurations
with the clover action pushing the light quark masses to
near the physical region\cite{pacscs}. As of this writing, these 
configurations are not available for our use.
Other fermion formulations could be considered, such as
staggered fermions\cite{stag}, domain-wall fermions\cite{dwf} 
or twisted mass QCD\cite{tmqcd} which presently
operate at u/d to s-quark mass ratios as small as 0.1, 0.217 and 0.16
respectively. However, how to combine NRQCD with
such approaches may require some attention. Also, the possibility of a
hybrid calculation where sea and valence quarks are treated using
different actions needs further consideration.

In our previous papers charmed baryons were also studied\cite{NMnrqcd,charm} 
and one would like to test one's capability of doing calculations in this
sector. At the lattice spacing used here the charm quark is too light 
for NRQCD so the treatment of charmed baryons is left for future work.

\appendix

\renewcommand{\theequation}{A\arabic{equation}}
\setcounter{equation}{0}

\section*{APPENDIX}

The heavy quark action is described using NRQCD\cite{latnrqcd}. 
The heavy quark propagator is given by
\begin{eqnarray}\label{HQprop}
\nonumber
G_{\tau+1} &=& \left(1-\frac{aH_B}{2}\right)
    \left(1-\frac{aH_A}{2n}\right)^n\frac{U_4^\dagger}{U_0} \\
           && {\times}\left(1-\frac{aH_A}{2n}\right)^n
    \left(1-\frac{aH_B}{2}\right) G_\tau,
\end{eqnarray}
with $n=5$ used in this work.
The Hamiltonian is separated into two terms, $H=H_A+H_B$, with $H_A$ containing the
kinetic piece $H_0$ and the term proportional to $c_{10}$ (defined below).

The Hamiltonian
contains all terms up to $O(1/M_0^3)$ in the classical continuum limit:
\begin{eqnarray}
H &=& H_0 + \delta{H}, \label{H} \\
H_0 &=& \frac{-\Delta^{(2)}}{2M_0}, \\
\delta{H} &=& \delta{H}^{(1)} + \delta{H}^{(2)} + \delta{H}^{(3)} + O(1/M_0^4), \\
\delta{H}^{(1)} &=& -\frac{c_4}{U_0^4}\frac{g}{2M}\mbox{{\boldmath$\sigma$}}
                    \cdot\tilde{\bf B} + c_5\frac{a^2\Delta^{(4)}}{24M_0}, \\
\nonumber
\delta{H}^{(2)} &=& \frac{c_2}{U_0^4}\frac{ig}{8M_0^2}(\tilde{\bf \Delta}
                    \cdot\tilde{\bf E}-\tilde{\bf E}\cdot\tilde{\bf \Delta})
                  -\frac{c_3}{U_0^4}\frac{g}{8M_0^2} \\
               && {\times} \mbox{{\boldmath$\sigma$}}
               \cdot(\tilde{\bf \Delta}\times\tilde{\bf E}-\tilde{\bf E}\times
               \tilde{\bf \Delta}) - c_6\frac{a(\Delta^{(2)})^2}{16nM_0^2}, \\
\nonumber
\delta{H}^{(3)} &=& -c_1\frac{(\Delta^{(2)})^2}{8M_0^3}
                    -\frac{c_7}{U_0^4}\frac{g}{8M_0^3}\left\{\tilde\Delta^{(2)},
                        \mbox{{\boldmath$\sigma$}}\cdot\tilde{\bf B}\right\} \\
\nonumber
                &&  -\frac{c_9ig^2}{8M_0^3}\mbox{{\boldmath$\sigma$}}\cdot
                     \left(\frac{\tilde{\bf E}\times\tilde{\bf E}}{U_0^8}
                    +\frac{\tilde{\bf B}\times\tilde{\bf B}}{U_0^8}\right)
                     \nonumber \\
             && -\frac{c_{10}g^2}{8M_0^3}\left(\frac{\tilde{\bf E}^2}{U_0^8}
                +\frac{\tilde{\bf B}^2}{U_0^8}\right)
                -c_{11}\frac{a^2(\Delta^{(2)})^3}{192n^2M_0^3}.
              \label{dH3}
\end{eqnarray}
The tildes indicate that the leading
discretization errors have been removed. In particular,
\begin{eqnarray}
   \tilde{E}_i &=& \tilde{F}_{4i}, \\
   \tilde{B}_i &=& \frac{1}{2}\epsilon_{ijk}\tilde{F}_{jk},
\end{eqnarray}
where
\begin{eqnarray}
\nonumber
   \tilde{F}_{\mu\nu}(x) &=& \frac{5}{3}F_{\mu\nu}(x) \nonumber
        - \frac{1}{6U_0^2}[U_\mu(x)F_{\mu\nu}(x+\hat\mu)U_\mu^\dagger(x) \\
\nonumber
   && +U_\mu^\dagger(x-\hat\mu)F_{\mu\nu}(x-\hat\mu)U_\mu(x-\hat\mu) \\
   && -(\mu\leftrightarrow\nu)]+\frac{1}{3}\left(\frac{1}{U_0^2}-1\right)F_{\mu\nu}(x).
\end{eqnarray}
The last term in $\tilde{F}_{\mu\nu}(x)$ corrects for the fact that the
gauge field link multiplied by a tadpole factor is no longer 
unitary\cite{groote}.

The spatial lattice derivatives are given by
\begin{eqnarray}
\nonumber
   a\Delta_iG(x) &=& \frac{1}{2U_0}[U_i(x)G(x+\hat\imath)
                    -U^\dagger_i(x-\hat\imath)G(x-\hat\imath)], \\
                    && \\
   a\Delta^{(+)}_iG(x) &=& \frac{U_i(x)}{U_0}G(x+\hat\imath) - G(x), \\
   a\Delta^{(-)}_iG(x) &=& G(x) -
              \frac{U^\dagger_i(x-\hat\imath)}{U_0}G(x-\hat\imath), \\
\nonumber
   a^2\Delta^{(2)}_iG(x) &=& \frac{U_i(x)}{U_0}G(x+\hat\imath) - 2G(x) \\
            &&  +\frac{U^\dagger_i(x-\hat\imath)}{U_0}G(x-\hat\imath), \\
   \tilde\Delta_i &=& \Delta_i
                   - {a^2\over 6} \Delta^{(+)}_i\Delta_i\Delta^{(-)}_i, \\
   \Delta^{(2)} &=& \sum_i \Delta^{(2)}_i \label{Laplacian}, \\
   \tilde \Delta^{(2)} &=& \Delta^{(2)} - {a^2 \over 12} \Delta^{(4)}, \\
   \Delta^{(4)} &=& \sum_i \left( \Delta^{(2)}_i \right)^2.
\end{eqnarray}

\acknowledgments
We thank the CP-PACS/JLQCD Collaborations for making their dynamical gauge 
field configurations available. Also, we thank Robert Petry for maintaining 
the VXRACK cluster at the University of Regina where our computations were 
done and Wendy Taylor for a discussion of the D0 Collaboration's 
$\Omega_b$ result. This work was supported in part by the 
Natural Sciences and Engineering Research Council of Canada.


\end{document}